\documentclass{article}
\usepackage{spconf,amsmath,graphicx}

\title{Effective internal language model training and fusion for factorized transducer model}

\name{\begin{tabular}{c}Jinxi Guo, Niko Moritz, Yingyi Ma, Frank Seide, Chunyang Wu, \\
Jay Mahadeokar, Ozlem Kalinli, Christian Fuegen, Mike Seltzer\end{tabular}}
\address{Meta AI}
\begin{document}
\ninept
\maketitle
\begin{abstract}
The internal language model (ILM) of the neural transducer has been widely studied. In most prior work, it is mainly used for estimating the ILM score and is subsequently subtracted during inference to facilitate improved integration with external language models. Recently, various of factorized transducer models have been proposed, which explicitly embrace a standalone internal language model for non-blank token prediction. However, even with the adoption of factorized transducer models, limited improvement has been observed compared to shallow fusion. In this paper, we propose a novel ILM training and decoding strategy for factorized transducer models, which effectively combines the blank, acoustic and ILM scores. Our experiments show a 17\% relative improvement over the standard decoding method when utilizing a well-trained ILM and the proposed decoding strategy on LibriSpeech datasets. Furthermore, when compared to a strong RNN-T baseline enhanced with external LM fusion, the proposed model yields a 5.5\% relative improvement on general-sets and an 8.9\% WER reduction for rare words. The proposed model can achieve superior performance without relying on external language models, rendering it highly efficient for production use-cases. To further improve the performance, we propose a novel and memory-efficient ILM-fusion-aware minimum word error rate (MWER) training method which improves ILM integration significantly.
\end{abstract}

\section{Introduction}
\label{sec:intro}

End-to-end models for automatic speech recognition (ASR) have gained increasing popularity in recent years as a way to fold separate components of a conventional ASR system (i.e., acoustic, pronunciation and language models) into a single neural network. Examples of such models include connectionist temporal classification (CTC) based models \cite{graves2006connectionist}, recurrent neural network transducer (RNN-T) \cite{graves2012sequence}, and attention-based seq2seq models \cite{chan2016listen}. Among these models, RNN-T is the most suitable streaming end-to-end recognizer, which has shown superior performance compared to others \cite{he2019streaming}.

However, the end-to-end architecture of RNN-T models introduces certain challenges. Without distinct acoustic and language components, it's not straightforward to train the internal language model with text-only data, which could potentially enhance the model capabilities. Therefore, many existing approaches utilize external LMs to improve the model performance. In~\cite{Hannun14}, a shallow fusion approach has been proposed and it combines the E2E model score with the LM score during beam search. The density ratio method~\cite{McDermott19} enhanced this by subtracting a source-domain LM score. A recent hybrid autoregressive model (HAT)~\cite{Variani20HAT} estimated and subtracted the internal LM score at the shallow fusion stage, notably improving ASR performance. \cite{Meng21ILM,Zeyer21} follow the similar ideas and apply it to standard RNN-T models.

Recently several factorized transducer (FT) models have been proposed. \cite{Xie22FT} first proposes a factorized neural transducer model, which factorizes the blank and non-blank predictor, in order to make the non-blank predictor behave as a standalone language model. In \cite{Meng22MHAT} a similar model architecture named as modular HAT is proposed, which applies a separate normalization over non-blank logits. While these related works indicate that FT models exhibit superior performance in domain adaptation tasks, their improvement on general datasets remains somewhat constrained when compared to shallow fusion approaches \cite{Meng23JEIT,Zhao23FT}.

In this work, we propose a novel training and decoding strategy for FT models, which effectively combine the blank, acoustic model (AM) and ILM scores. The proposed method first trains the ILM with text-only data, and then jointly optimizes the FT model with RNN-T loss. At the decoding stage, a novel ILM fusion strategy has been proposed, which is controlled by two ILM weights applied on both inside and outside of the non-blank score calculation. Once the optimized weights are selected, we apply a novel ILM-fusion-aware MWER training on top of the FT model and the decoding set-up, to further improve the ILM integration. To reduce the memory usage of MWER training, an alignment-restricted MWER training method has been proposed. This method leverages the estimated maximum-probability alignment obtained from beam search decoding, in conjunction with predefined left and right context, to restrict the alignment summation paths used for RNN-T probability calculations.
\raggedbottom

\section{Neural Transducer Model}
\label{sec:Neural Transducer Model}

\subsection{RNN-T Model}
An RNN-T model consists of an encoder, a prediction network and a joint network. The encoder receives acoustic feature vectors $\mathbf{x}_t$ and converts them into a sequence of hidden states $\mathbf{h}_{t}^{enc}$, where $t$ is the time index. The prediction network takes previous sub-word label prediction $\mathbf{y}_{u-1}$ as input, and produces hidden representation $\mathbf{h}_{u}^{pre}$, where u is label index. The joint network is a feed-forward network that takes the combination of encoder output $\mathbf{h}_{t}^{enc}$ and prediction network output $\mathbf{h}_{u}^{pre}$, and computes output logits $\mathbf{h}_{t,u}$. The final posterior for the output token $P_{t,u}$ is obtained after applying the softmax operation.

The loss function of RNN-T is the negative log posterior of output label sequence $\mathbf{y}$ given input acoustic feature $\mathbf{x}$: 
\begin{equation}
L_\text{RNN-T} = -\log{P(\mathbf{y}|\mathbf{x})},
\label{eq:1}
\end{equation}
where $P(\mathbf{y}|\mathbf{x}) = \sum_{\hat{\mathbf{y}}}P(\hat{\mathbf{y}}|\mathbf{x})$, $\hat{\mathbf{y}} \in A$. $A$ is the set of all the possible alignments (containing both blank and non-blank labels) between input $\mathbf{x}$ and output $\mathbf{y}$. 

\subsection{Factorized Transducer Model}
\begin{figure}[t]
\centering
\includegraphics[width=0.95\linewidth]{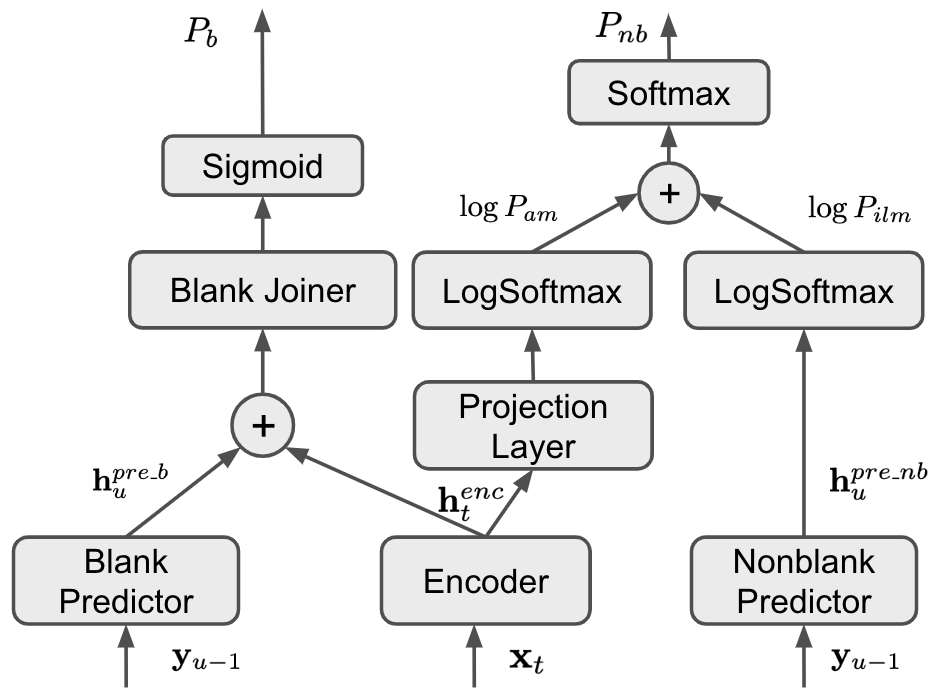}
\caption{Factorized predictor model architecture.}
\label{fig:FT}
\end{figure}

Standard RNN-T models use a shared predictor and a shared joiner to predict the joint posterior distribution of both blank and non-blank tokens. Recently, several factorized transducer (FT) models have been proposed~\cite{Xie22FT,Meng22MHAT}, which use separate predictors and joiners for blank and non-blank token prediction. In this paper, we use a very similar factorized predictor model architecture for FT. As shown in Figure \ref{fig:FT}, for the factorized predictor model, we have 2 separate predictors, i.e. a blank predictor and a non-blank predictor. They both take $\mathbf{y}_{u-1}$ as input, and generate hidden representation $\mathbf{h}_{u}^{pre\_b}$ and $\mathbf{h}_{u}^{pre\_nb}$. The encoder receives acoustic feature vectors $\mathbf{x}_t$ and convert it to $\mathbf{h}_{t}^{enc}$. The $\mathbf{h}_{t}^{enc}$ is shared between the blank branch (left side) and the non-blank branch (right side). For the blank branch, $\mathbf{h}_{u}^{pre\_b}$ and $\mathbf{h}_{t}^{enc}$ are added together, and then pass to a blank joiner model to generate a single-dimension logit. The logit is then passed to a Sigmoid function to generate the blank probability $P_{b}$. For the non-blank branch, the acoustic hidden representation $\mathbf{h}_{t}^{enc}$ is first passed to a projection layer, in order to match the output vocabulary size. Then a LogSoftmax function is used to convert the projection output into a log scale acoustic probability ($\log{P_{am}}$). Similarly, we also apply a LogSoftmax function on top of the non-blank predictor hidden representation $\mathbf{h}_{u}^{pre\_nb}$, and generate the log-scale internal language model probability ($\log{P_{ilm}}$). The $\log{P_{am}}$ and $\log{P_{ilm}}$ scores are then added together, and further normalized using a Softmax function to generate the non-blank probability $P_{nb}$. To get the final posterior, we use the same method as in \cite{Variani20HAT} and normalize the posterior as:
\begin{equation}
\begin{cases}
P_{b}& \text{if}\:\text{token = blank}\\
(1-P_{b})P_{nb} & \text{if}\:\text{token = non\_blank}
\end{cases}
\label{eq:2}
\end{equation}

\section{ILM training and fusion}
\label{sec:ILM training and fusion}

\subsection{ILM Training}

ILM training optimizes an additional ILM loss together with the RNN-T loss. For standard RNN-T and HAT models, the ILM loss could use the ILM score estimated from the output of the predictor and joiner excluding the blank and acoustic connection \cite{Variani20HAT, Meng21ILM}. For the factorized transducer model used in this paper, we can get an explicit ILM probability directly from the output of the non-blank predictor and the LogSoftmax function, i.e. $\log{P_{ilm}}$. 
Therefore the ILM loss is defined as below:
\begin{equation}
L_\text{ILM} = -log{P_{ilm}}.
\label{eq:3}
\end{equation}
ILM training can help to get accurate ILM probabilities, and thus improve the fusion with external LMs. More importantly, ILM training could help the model to learn from the text-only data, which may vary significantly or be substantially larger than the training data transcript. ILM loss can be optimized separately or jointly with the RNN-T loss, and both approaches can help to learn a good standalone ILM model. In this paper, we investigate the approach of first pre-training the non-blank predictor with large-scale text-only data using ILM loss, and then fix the non-blank predictor for RNN-T loss training. 

\subsection{Proposed ILM fusion strategy}
\label{sec:Proposed ILM fusion strategy}
Previous research on factorized transducer models uses the same decoding strategy as RNN-T and HAT models, i.e. calculating the blank and non-blank scores same as what is defined at training stage as shown in Eq.~\ref{eq:4}
\begin{equation}
\begin{cases}
\log P_{b}& \text{blank}\\
\log ((1-P_{b})(\text{softmax}(\log P_{am}+\log P_{ilm}))) & \text{non\_blank}
\end{cases}
\label{eq:4}
\end{equation}

Our experimental results show that, even though a well-trained non-blank predictor using the ILM loss and large-scale text-only data can help to generate a better $\log{P_{ilm}}$ with low perplexity, the improvement of this ILM training is not significant with the existing decoding strategy. Therefore, in this paper, we propose a new decoding strategy to modify the score of non-blank token as shown in Ep.\ref{eq:5}. 
\begin{equation}
\log ((1-P_{b})(\text{softmax}(\log P_{am}+\alpha*\log P_{ilm})))+\beta*\log P_{ilm}
\label{eq:5}
\end{equation}
Inspired by the traditional Bayes rule of combining AM and LM scores and also shallow fusion, the same ILM score $\log{P_{ilm}}$ is added on top of the existing non-blank score, with a weight $\beta$ to scale the log probability. Moreover, inspired by the internal language model subtraction method~\cite{Variani20HAT}, a weight $\alpha$ is applied on top of the non-blank predictor and the LogSoftmax function output. $\alpha$ ranges from 0-1 at the decoding stage, and can be used to down-scale and subtract the ILM, when combining with other scores. We will show in Section \ref{sec:Results} that, both weight $\alpha$ and $\beta$ are critical to achieve a superior performance of this factorized transducer model.

\subsection{Proposed ILM-fusion-aware and memory-efficient MWER training}

Sequence discriminative training has shown to be very effective for both hybrid \cite{vesely2013sequence} and end-to-end ASR models~\cite{graves2014towards,Shannon17, Rohit18, Jinxi20MWER}, and it can also help to optimize the fusion between AM/end-to-end models and LM models with a unified training loss \cite{vesely2013sequence, Zhong21MWERLM, Weiran21MWERLM}. Since in Section~\ref{sec:Proposed ILM fusion strategy} a new ILM score fusion strategy has been proposed, it's nature to propose an ILM-fusion-aware sequence discriminative training approach to further enhance the optimization of ILM fusion. 

We first generate the N-best hypotheses $y_i$ for each batch, using the decoding set-up in Eq.~\ref{eq:5}. Following the same method proposed in \cite{Jinxi20MWER}, we calculate the minimum word error rate (MWER) loss with the full-sum alignment probabilities from RNN-T loss, and the proposed ILM probabilities is also added as shown in Eq.~\ref{eq:6}. We denote the number of word errors in $y_i$ by $R(y_{i},y^{r})$, where $y^r$ is the reference. For simplicity, the same $\alpha$ and $\beta$ weights from Section.~\ref{sec:Proposed ILM fusion strategy} are used in Eq.\ref{eq:6} during training to generate the final probability scores of each hypotheses.
\begin{eqnarray}
\begin{split}
L_\text{MWER} = & \sum_{y_{i} \in \text{nbest}(x)}\hat{P}(y_{i}|x)R(y_{i},y^{r}), \\
\text{where}~\hat{P}(y_{i}|x) =~&\text{softmax}(\log{P(y_{i}|x)} + \beta*\log P_{ilm}) 
\end{split}
\label{eq:6}
\end{eqnarray}

We use the alignment-restricted RNN-T loss \cite{Jay21AR} for our factorized transducer training, which exhibits comparable WER performance while significantly improving GPU memory efficiency. Hence, for our proposed MWER training, we aim to implement a similar alignment restriction method for computing scores for each hypothesis, thereby effectively reducing memory usage. However, in order to apply alignment restriction in RNN-T loss, pre-computed alignments generated from a hybrid ASR alignment model are needed, which is impossible to get during MWER training. Therefore, in this paper, we proposed a novel method to estimate the alignments from the beam-search decoding.
More precisely, during beam-search decoding, we update and store the word alignments of each hypothesis with the highest probability. Upon obtaining the final alignment, we employ predefined left and right contexts for each timestamp, which adds a tolerance margin and is used to confine the number of alignments (or paths) within the RNN-T lattice. Only the alignments within the left and right context of the final alignment will be used for the RNN-T loss calculation.

In the realm of gradient calculation, we utilized the chain rule to back-propagate the MWER gradients to RNN-T losses. For RNN-T loss gradients, we implemented a memory-efficient forward-backward algorithm by only considering the forward and backward variables within the alignment-restriction paths mentioned above. The adoption of this method significantly reduces the memory usage associated with MWER training. Moreover, to maintain the model's token emission latency, we incorporate the RNN-T loss (utilizing ground-truth alignments) with a small weighting alongside the proposed ILM-fusion-aware MWER loss.

\section{Experimental Setup}
\label{sec:experimental_setup}

\subsection{Training and evaluation data sets}

We conduct our experiments on LibriSpeech data sets~\cite{Libri15}. The training data contains around 960 hours of speech from read audio book recordings. SpecAugment \cite{park2019specaugment} is applied on-the-fly during training. WERs and Rare Word (RW) WERs are reported on the test-clean and test-other splits. As for the text-only training set to train ILM and external LM, we use the 800M word LibriSpeech language modeling corpus.

\subsection{Models and Training Setup}
For the baseline RNN-T model, the encoder is a 20-layer streamable low-latency Emformer \cite{Yangyang21emformer}, 40 ms lookahead, 160 ms segments, input dimension 512, hidden dimension 2048, 8 self-attention heads, and 1024-dimensional FC projection. The predictor consists of three LSTM layers with 512-dim hidden size. The output vocabulary is 5000 unigram SentencePieces estimated from the training transcripts, plus the blank symbol. For the FT model, the same encoder model is used. For the blank predictor, a stateless predictor is used with an embedding layer and a linear projection layer. For the non-blank predictor, 2 different set-ups have been explored: one set-up uses the same predictor model as the RNN-T baseline, and the second one employs a larger-size predictor featuring two LSTM layers with hidden dimensions of 2048 and projection dimensions of 512. We also train an external language model on the text-only LM corpus with the same model architecture as the second non-blank predictor, in order to perform shallow fusion with the baseline RNN-T model.

We train all models using the alignment-restricted RNN-T loss, where the alignment is provided by a hybrid acoustic model. For baseline RNN-T model, we jointly train RNN-T loss with the ILM loss as in \cite{Meng21ILM}, and perform ILM subtraction when doing shallow fusion with the external LM. For FT model, the non-blank predictor is first pre-trained with the large-scale text-only LM corpus, and then fixed for RNN-T loss training. For the ILM-fusion-aware MWER training, we use a decoding beam size of 5, and the ILM is fixed during training.

\section{Results}
\label{sec:Results}

\subsection{Baseline RNN-T model and factorized transducer model}

\begin{table}[t]
  \centering
  \begin{center}
    \begin{tabular}{ | l || c | c | c  | c  |p{5cm}|}
    \hline
    Models & Clean & RW & Other & RW \\ \hline
    RNN-T 77M  & 3.52  & 10.15 & 8.84 & 23.4   \\ \hline
    + LM SF 100M & 3.01 & 8.61 & 7.72 & 20.59   \\ \hline
    FT 77M  & 3.78 & 10.29 & 9.12 & 24.1   \\ \hline
    FT 95M  & 5.79 & 16.98 & 11.53 & 30.65  \\ \hline
    + ILM pre-trained  & 3.50 & 9.12 & 8.78 & 22.88  \\ \hline
    \end{tabular}
  \end{center}
  \caption{Baseline WER results for RNN-T and factorized transducer (FT) models. RW denotes rare words.}
  \label{table:baseline}
\end{table}

We first compare the model performance between the baseline RNN-T model and FT model using the standard decoding strategy. Given a similar number of parameters (i.e 77M), FT performs a bit worse compared to the RNN-T model. That's likely due to the fact that LibriSpeech training set is relatively small, and FT model needs more training data to reach a better performance. When we increase the non-blank predictor size (i.e 95M FT model), the model starts to overfit on the training set and thus show relatively poor generalization results. Therefore, we pre-train the non-blank predictor with large-scale text-only data, and then fix it for RNN-T loss training. As shown in the last row of Table~\ref{table:baseline}, 95M FT model with ILM pre-trained, starts to show better performance compared with the 77M FT model, especially on Rare Word (RW) WERs. However, the FT model performance is still just slightly better than RNN-T 77M baseline model, and significantly worse than RNN-T + external LM shallow fusion (with optimized ILM weight -0.2, external LM weight 0.6).  

\subsection{Performance of the proposed ILM fusion strategy}

To improve the performance and decoding strategy of the FT model, we first try to add an additional ILM score with a weight $\beta$ on top of the existing non-blank score. As shown in Table~\ref{table:proposed ILM fusion strategy} (row 5-7), the WERs improve significantly across all test sets, and 0.2 gives the best performance.
A natural question arises: can we simply increase the ILM weight $\alpha$ when combining it with the AM score $\log{P_{am}}$ (for training, $\alpha$ is 1)? The results in row 9 indicate that, only increasing the weight $\alpha$ doesn't improve the model performance. This could be because that, adding ILM score externally instead of increasing ILM weight $\alpha$ will not impact the learned relationships between AM, ILM and blank scores, but provide useful LM information for non-blank tokens prediction.

For RNN-T and HAT model, the ILM scores are estimated by zeroing out the contributions of the acoustic encoder and they are applied at the shallow fusion stage.
For FT models, this is unnecessary, as the non-blank predictor already furnishes an explicit ILM score. Therefore, we can perform ILM subtraction by simply decreasing the weight $\alpha$, instead of subtracting it at the shallow fusion step (which is less accurate). As shown in row 8, simply decreasing $\alpha$ will degrade the model performance, and therefore we also add an ILM score externally with a tuned value for $\beta$.
Finally, we find that $\beta=0.6$ and $\alpha=0.6$ give the best performance, which is 17\% better compared to the standard FT model. It also show superior performance compared to the baseline RNN-T + shallow fusion, especially on Rare Words (8.9\% relative improvement). The proposed method offers a solution for maximizing the utility of a well-trained ILM model while circumventing the need for an additional external LM, which could otherwise lead to increased computation and memory usage.

\begin{table}[t]
  \centering
  \begin{center}
    \begin{tabular}{ | l || c | c | c  | c  |p{5cm}|}
    \hline
    Models & Clean & RW & Other & RW \\ \hline
    RNN-T & 3.52  & 10.15 & 8.84 & 23.4   \\ \hline
    + LM SF (0.6, -0.2) & 3.01 & 8.61 & 7.72 & 20.59   \\ \hline
    FT + ILM pre-trained  & 3.50 & 9.12 & 8.78 & 22.88  \\ \hline
    + $\beta=0.1$, $\alpha=1$ & 3.19 & --- & 8.08 & ---  \\ \hline
    + $\beta=0.2$, $\alpha=1$ & 3.06 & 8.14 & 7.73 & 20.34  \\ \hline
    + $\beta=0.6$, $\alpha=1$ & 3.17 & --- & 7.98 & ---  \\ \hline
    + $\beta=0$, $\alpha=0.6$ & 3.88 & --- & 9.78 & ---  \\ \hline
    + $\beta=0$, $\alpha=1.3$ & 3.52 & 9.35 & 8.74 & 22.69  \\ \hline
    + $\beta=0.4$, $\alpha=0$ & 5.12 & 16.26 & --- & ---  \\ \hline
    + $\beta=0.8$, $\alpha=0$ & 4.77 & 14.57 & --- & ---  \\ \hline
    + $\beta=0.6$, $\alpha=0.6$ & 2.89 & 7.84 & 7.29 & 19.29  \\ \hline
    - length normalization & 6.07 & 11.97 & 15.53 & 28.29  \\ \hline
    + ILM-aware MWER & 2.87 & 7.68 & 7.31 & 18.90  \\ \hline
    \end{tabular}
  \end{center}
  \caption{WER numbers of applying the proposed ILM fusion strategy and ILM-fusion-aware MWER training.}
  \label{table:proposed ILM fusion strategy}
\end{table}

\subsection{Effect of the proposed ILM-fusion-aware MWER training}

The weights $\alpha$ and $\beta$ are manually tuned in Section 5.2, and ILM-fusion-aware MWER training could help to incorporate those decoding parameters and strategies into model optimization. Therefore, we use the proposed decoding set-up to generate N-best and calculate hypotheses scores. Please note that, the hypothesis scores in Eq.~\ref{eq:6} employed in MWER training do not use length normalization. Consequently, we also turn off length normalization for N-best generation. While model performance degrades significantly without length normalization, the proposed ILM-fusion-aware MWER training effectively enhances the WER performance, and demonstrates further improvement in Rare Word WERs (10.8\% relative improvement). Based on our recent large-scale ASR training experiments, the proposed MWER training approach shows a more substantial improvement when trained on larger datasets.

We further experiment with the proposed alignment-restricted MWER loss using different left and right contexts. Using more contexts increases memory consumption, so we seek a balance in selecting left and right contexts that can cover most high-probability alignments. In the end, we found that using a left and right context of 15 frames can give us reasonable performance improvement, while keeping the GPU memory to be significantly smaller.

\section{Conclusion}
\label{sec:Conclusion}

In this paper, we proposed a novel internal language model (ILM) training and decoding strategy for factorized transducer models. The proposed method effectively combines the blank, acoustic, and ILM scores, leading to a substantial performance improvement.
Our experiments demonstrated a 17\% relative improvement when utilizing a pre-trained ILM and the proposed decoding strategy for factorized transducer models on LibriSpeech datasets.
Furthermore, when compared to a strong RNN-T baseline with external LM shallow fusion, the proposed model yielded about 5.5\% relative improvement on the general test sets of LibriSpeech and a 8.9\% WER reduction for rare words. The proposed model can achieve superior performance without requiring an additional external language model, making it highly efficient for production use-cases.
To further improve the performance, we proposed a novel and memory-efficient ILM-fusion-aware MWER training method which significantly improves ILM integration. A more substantial improvement has been observed based our latest large-scale ASR training experiments.
In conclusion, our work provides a new perspective on the training and decoding strategy for factorized transducer models, and opens up new possibilities for further research in this area.

\bibliographystyle{IEEEbib}
\bibliography{strings,refs}

\end{document}